\documentclass[aip,jcp,reprint]{revtex4-1}
\usepackage{fullpage}
\usepackage[latin1]{inputenc}
\usepackage{amsmath,esint}
\usepackage{amsfonts}
\usepackage{amssymb}
\usepackage{color}
\usepackage{graphicx, graphics}
\usepackage{epsfig}
\usepackage{subfig}
\usepackage{wasysym}
\begin{document}

\title{Hot Nano-particles in Polar or Paramagnetic Liquids Interact as Monopoles. }
\author{Daan Frenkel$^*$}
\affiliation{ Department of Chemistry, University of Cambridge, Lensfield Road, Cambridge CB2 1EW, UK }
\email{df246@cam.ac.uk}
\begin{abstract} 
When neutral nano-particles are heated or cooled in a polar liquid, they will interact with each other as if they carry an electrostatic charge that is proportional to the temperature difference between the particle and the surrounding fluid. The same should hold for suspensions liquids of asymmetric ferromagnetic particles, in which case the heated nano-particles should behave as magnetic monopoles.  However, the analogy with electrostatics/magnetostatics is not complete: heated/cooled nano-particles do not move under the influence of an applied homogeneous field. They should, however, interact as monopoles with each other and should move in inhomogeneous fields. 
\end{abstract}
\maketitle
\section{Introduction}
Electrostatics and magneto-statics are often considered to be simple subjects, but they are not.
As an illustration I consider the curious effect that results when neutral, colloidal particles in a polar (or paramagnetic) liquid are heated or cooled with respect to the surrounding fluid (for the sake of brevity, I will henceforth write `heated' instead of `heated or cooled').  I will argue that these particles then interact as if they carry a charge (or, in the case of a paramagnetic solvent, as if they carry a magnetic monopole). This charge is unscreened (except in the presence of added salt). However, as I will argue below, moving the monopole against a homogeneous external field requires no work and, conversely, an external field cannot do work on the monopole.  Hence, homogeneous external fields cannot create currents of such monopoles.

To understand how heated neutral nano-particles can emulate charge, we note two things: first of all, in a thermal gradient, a polar liquid of particles with a sufficiently low-symmetry shape  will be polarised. As an example, water is known to show this effect (at least in simulations)~\cite{Bresme2008}. In what follows, I consider electrical polarisation. However, the argument for magnetic monopoles is basically the same. To lowest order, the resulting polarisation ${\bf P}$ is linear in the temperature gradient:
\[
{\bf P}= a{\bf \nabla} T
\]
where $a$ is a constant that depends on the thermodynamic state and the molecular details of the liquid. In a suspension of uncharged nano-particles, there are no free charges and hence the dielectric displacement ${\bf D}$ vanishes. 

Using ${\bf D}=\epsilon_0{\bf E} +{\bf P} = 0$, it follows that, around a nano-particle,
\[
{\bf E}= - {\bf P}/\epsilon_0
\]
where $\epsilon_0$ is the dielectric permittivity of vacuum. Using the fact that ${\bf P}$ is linear in ${\bf \nabla} T$, we obtain
\[
{\bf E}= -\alpha {\bf \nabla} T
\]
where we have defined $\alpha\equiv a/\epsilon_0$. For water near room temperature, $\alpha$ has been estimated to be $\alpha\approx$ 0.5 mV/K~\cite{Bresme2015,Wirnsberger2016}. 

The next thing that we need is the equation for heat diffusion. For convenience, I will assume that we are in the linear response regime. In the linear regime, the temperature in the liquid obeys the Laplace equation $\nabla^2 T$=0. Suppose that the temperature just outside the surface of a nano-particle of radius $R$ is $T_R$ and the temperature in the bulk of the liquid is $T_\infty$, then, in steady state,  the temperature profile at a distance $r$ from an isolated, spherical nano-particle satisfies:
\[
T(r)= T_\infty +(T_R-T_\infty){R\over r}
\]
and hence
\[
{\bf E}(r) = \alpha (T_R-T_\infty){R\over r^2} {\hat r}
\]
Note that $E$ decays as $1/r^2$. Using Gauss's theorem, we can then write
\[
\oiint {\bf E}(r) \cdot d{\bf S} = 4\pi\alpha (T_R-T_\infty)R \equiv q_{eff}/\epsilon_0 \;.
\]
In words: the flux through a closed surface around a neutral nano-particle is non-zero, it is equal to the flux due to an apparent charge $q_{eff} = 4\pi a (T_R-T_\infty)R$. Note that the effective charge is proportional to the radius of the particle, hence larger colloids will have a larger apparent charge. 

Of course, the existence of apparent monopoles in condensed matter systems has been established in other systems, e.g. in  spin ice~\cite{Castelnovo}. However, the current example does not rely on a particular lattice ordering. Also,  temperature induced monopoles need not occur in pairs of opposing charge.  Finally, as I will argue below, heat-induced monopoles couple differently  with external fields.

As a consequence of the temperature-induced fields,  a hot nano-particle repels another hot particle, but attracts a particle that is cooled below the temperature of the ambient liquid.  I will argue that this effect is real, but it may be hard to observe because dissolved nano-particles in thermal gradient will also be subject to thermo-phoretic forces. Separating the effects, whilst not impossible, will not be easy. The main reason why the effects should be separable is that thermo-phoretic forces are (in the linear regime) a  linear superposition of the phoretic force of heated particle 1 on unheated particle 2, and of heated particle 2 on unheated particle 1. In contrast, the `electrostatic' effect depends on the product of the degrees of heating of the two particles. 

We can quickly establish two results: 1)  that no work is needed to move a heated nano-particle in a uniform electric field. Conversely, this means that a uniform electric field does not drive the motion of heated or cooled nano-particles. And 2) that two heated nano-particles should experience electrostatic repulsion

Let us first consider a heated nano-particle in a uniform field ${\bf E}_0$. The electrostatic energy of this particle has two contributions: first of all, the electrostatic self-energy of the heated particle (i.e. the energy of the electrostatic field around an isolated, heated nano-particle). This self energy does not depend on position. The second contribution is due to integral $-\int d{\bf r}\; {\bf E}_0\cdot {\bf P}({\bf r})$. This integral vanishes for any position of the nano-particle. Hence, moving a heated nano-particle in a uniform electric field requires no work - hence, the field can do no work on the particle and it cannot, therefore, drive a DC current. However, other fields could drive thermal monopoles. For instance, a single thermal monopole might move under the influence of a homogeneous concentration gradient (diffusiophoresis) or a thermal gradient (thermophoresis). However, when the monopoles reach the walls of the container, they must stop, as the charge cannot be transferred to an electrode. 

It is interesting to compare the behaviour of thermal monopoles with that of magnetic monopoles in spin ice~\cite{Castelnovo}: when spin-ice monopoles move, they flip spins and hence change the total magnetisation of the system, thus changing the energy of the system in a homogeneous external field. As a consequence, an applied magnetic field can induce a transient current of monopoles (transient, because the bulk polarisation is bounded). 

Next consider two thermal monopoles. We will assume that one particle is already heated and is therefore surrounded by a radially symmetric E-field that decays as $1/r^2$. Initially, the other particle is at ambient temperature and is therefore not affected by the $E$-field due to the first particle. If we now start heating the second nano-particle, we create an additional  (spherically symmetric) polarisation around it. As before, we ignore the (position-independent) self-energy associated with this polarisation. However, as we polarise the second particle, we do electrical work against the field $E_1(r)$ of the first particle:
 \[
 \delta w = -\int d{\bf r}\; {\bf E}_1(r)\cdot \delta{\bf P}_2({\bf r})
\]
Strictly speaking, this is the work that would have to be performed if the same polarisation were to be created by mechanical means, rather than by a heat flux. 
As $P$ and $E$ are proportional, the net effect is that the work that we have to perform to bring two heated nano-particles to a distance $r$ is the same as the electrostatic energy of interaction between two nano-particle with charges $q_{eff}$ (in that case, too, we can write the total electrostatic energy as the sum of the two self-energies and a cross-energy. The latter has the same functional form as in the case of two heated nano-particles). 

Although heated nano-particles cannot be moved by applying a homogeneous external field, they do interact with particles (or rods) that carry real charge. This should facilitate the observation of the effect.
\section{Partial analogy with ions in polar liquids}
The idea that the apparent charge of a particle may be very different form its real charge is, of course, not new. A prototypical example is an ion with charge $q$ in a liquid with relative dielectric constant $\epsilon_r$. At a distance $r$, the magnitude of the  $E$-field due to this ion is equal to $E_u(r)/\epsilon_r$, where $E_u(r)$ is the field around the same ion in vacuum. We can interpret the reduction of the $E$ field due to the dielectric constant as being due to the original `bare' charge $q$ and a compensating `polarisation' charge $(1/\epsilon_r-1)q$. But, as in the case of heated nano-particles, an external electric field only couples to the bare charge, not to the polarisation charge. 
Also, when two ions in solution interact, the strength of the interaction is proportional to the product of the effective charge and the bare charge, unlike heated nano-particles, where the interaction is proportional to the product of the effective charges.
The easiest way to see why ions behave differently is to consider a charging process where one charged ion is already present in solution and a second ion at distance $r$ is being charged. The work needed for the charging contains two contributions: one is equal to the self energy which, as usual, we ignore because it does not depend on the position of the other ion. The other is the work needed to introduce a charge $q_2$ at a position where the electrostatic potential due to the first ion is $\phi(r)$.  Clearly, in the linear regime, this work is $\phi(r) q_2$. However, as $\phi(r)=q_1/(4\pi\epsilon_0\epsilon_r r)$, the total electrostatic energy is
\[
V_{el}(r) = {q_1q_2\over 4\pi\epsilon_0\epsilon_r r} \;,
\]
which should come as no surprise: the `screened' ion sees an `unscreened' ion. 

It is interesting to note that a volume heated in an electrolyte solution also can behave as a charged object, due to the temperature driven gradient in the concentration of the constituent anions and cations~\cite{Wurger}. For this effect to work, the Seebeck coefficient should be non-zero. It does not occur in liquids without free charge carriers and clearly has no magnetic analog. 
\section{Can it be measured?}
The effects described here should be observable, but the experiments may be challenging. In particular, as thermophoresis may obscure the predicted effects. However,  the effects can in principle be separated through their different dependence on the heating/cooling of the nano-particles.
Moreover, heated nano-particles will be attracted/repelled by a charged object that creates a field gradient (e.g. a charged wire). Changing the sign of the charge of the wire should deform the surrounding thermal nano-particle cloud. But, of course, for these experiments to work, it is crucial that the nano-particle is not only uncharged, but also carries not dipole moment. 

Another interesting question is how to make heated/cooled nano-particles. Heating is easy: it can be achieved by illumination. Cooling may be more difficult, although recent experiments have shown that laser cooling of colloidal particles is feasible~\cite{Pauzauskie}. Making a homogeneous mixture of hot and cold particles may be challenging. 

However, cooling particles may not be necessary to observe attraction: the alternative is to exploit an `image-charge' effect. In the presence of a flat wall that is both thermally and electrically conducting, a heated nano-particle should create an image-charge that mimics a cooled particle. 

A special problem is the creation of a fluid or suspension of low-symmetry molecular (or nano-colloidal) magnets. A direct coupling of the  polarity of the magnet with the molecular polarity is symmetry forbidden, as the magnetic moment has odd time-reversal symmetry, whereas all `structural' molecular properties such as polarity or chirality have even time-reversal symmetry. We can therefore at best create metastable molecular or colloidal magnets with broken up-down spin symmetry. Note that the magnetisation will have to be created while the particles are aligned, e.g. at an interface or, in the case of polar molecules, by an external electric field. Whilst aligned, the particles must be cooled below their blocking temperature, i.e. the temperature below which the magnetic moment of the particles is frozen in. For (large enough) low-symmetry nano-colloids consisting of a ferromagnetic material, it  is not a problem to reach the conditions where the magnetism is blocked. Permanent, polar molecular magnets are more challenging: at present, molecular magnets tend to have blocking temperatures well below room temperature -- but the values are increasing.

Nevertheless, in view of the serious experimental challenges, any initial test of the effect is likely to be numerical. 

\section*{Acknowledgments}
I gratefully acknowledge 4 decades of discussions with Bill Gelbart about science, espresso-making and cooking -- and any combination thereof. In addition, I gratefully acknowledge discussions with:  Paul Chaikin, Sharon Glotzer, Michael Brenner, David Buckingham, Ignacio Pagonabarraga, Lyd{\'e}ric Bocquet, Benjamin Rotenberg, Patrick Warren, Erika Eiser, Mike Cates, Ronojoy Adhikari, Christoph Dellago, Alpha Lee, Michiel Sprik and, in particular, Peter Palffy, Claudio Castelnovo, Mischa Bonn and Peter Wirnsberger. However, all mistakes are mine.

 \end{document}